\documentclass[%
 reprint,
 amsmath,amssymb,
 aps,
]{revtex4-2}

\usepackage{graphicx}% Include figure files
\usepackage{dcolumn}% Align table columns on decimal point
\usepackage{subfigure}
\usepackage[normalem]{ulem}
\usepackage{bm}% bold math
\begin{document}

\preprint{APS/123-QED}

\title{Anomalous transport of a classical wave-particle entity in a tilted potential}% Force line breaks with \\
%\thanks{A footnote to the article title}%

\author{Rahil N. Valani$^{1}$}\email{rahil.valani@adelaide.edu.au}

\affiliation{$^1$School of Mathematical Sciences, University of Adelaide, South Australia 5005, Australia} 

%\collaboration{CLEO Collaboration}%\noaffiliation

\date{\today}% It is always \today, today,
             %  but any date may be explicitly specified

\begin{abstract}

A classical wave-particle entity in the form of a millimetric walking droplet can emerge on the free surface of a vertically vibrating liquid bath. Such wave-particle entities have been shown to exhibit hydrodynamic analogs of quantum systems. Using an idealized theoretical model of this wave-particle entity in a tilted potential, we explore its transport behavior. The integro-differential equation of motion governing the dynamics of the wave-particle entity transforms to a Lorenz-like system of ordinary differential equations (ODEs) that drives the particle's velocity. Several anomalous transport regimes such as absolute negative mobility (ANM), differential negative mobility (DNM) and lock-in regions corresponding to force-independent mobility, are observed. These observations motivate experiments in the hydrodynamic walking-droplet system for the experimental realizations of anomalous transport phenomena.

\end{abstract}

\maketitle
%\tableofcontents
%    \item Introduce the phenomena of negative mobility by first starting with say non-nequilibrium systems: what they are and they different types of anamolous transport behaviours they exhibit. Highlight that one such behaviour is negative mobility and give different examples of the two types of negative mobility and the different systems in which they are observed. Maybe towards the end have the systems where negative mobility is observed in chaotic systems
\textit{Introduction.} When a biased force is applied to a particle having no net drift, one expects the particle to drift in the direction of the applied force. However, seemingly paradoxical behaviors have been observed in non-equilibrium systems that may result in anomalous response of the particle under applied bias. For example, absolute negative mobility (ANM) may arise where the particle responds with a net drift in a direction opposite to the applied bias. A less pronounced related phenomenon is differential negative mobility (DNM), where the particle drifts in the same direction as the applied bias but the drift speed of the particle decreases with increasing applied bias. ANM has been observed experimentally and theoretically in both quantum and classical systems~\citep{Eichhorn2005}. Examples of quantum systems that exhibit ANM include quantum-well structures~\citep{Hopfel1986} and semiconductor superlattices~\citep{Keay1995,Cannon2000}. In classical systems, ANM has been mainly observed and investigated in systems driven by noise such as single~\citep{cANMs1,cANMs2,cANMs3,cANMs4,cANMs5,cANMs6,cANMs7,cANMs8} and interacting~\citep{cANMm1,cANMm2,cANMm3,cANMm4,cANMm5} Brownian particles, while fewer works have studied ANM in deterministic systems. Examples of deterministic systems that exhibit ANM include vibrational motors~\citep{vibmotor}, a particle with space-\citep{spacedep} and speed-dependent damping~\citep{speeddep}, a particle in a time-varying potential~\citep{timevarypot}, a particle in a periodic double-well potential~\citep{doublewell} and a particle in a traveling wave system~\citep{travelwave}. Inspired by the emergence of ANM in deterministic systems, in this Letter we theoretically and numerically investigate the anomalous transport behavior of a self-propelled classical wave-particle entity in a tilted potential.

A liquid bath when vibrated vertically can support millimetric droplets on its free surface that walk horizontally while bouncing vertically~\cite{Couder2005,Couder2005WalkingDroplets,superwalker}. The walking droplet, also known as a walker, upon each bounce generates a damped localized standing wave on the fluid surface. It then interacts with these self-generated waves on subsequent bounces to propel itself horizontally. The droplet and its underlying wave field coexist as a wave-particle entity; the droplet generates the underlying wave field which in turn guides the motion of the droplet. At large vibration amplitudes, the waves created by a walker decay very slowly in time and the walker's motion is not only influenced by the wave created on its most recent bounce, but also by the waves generated in the distant past, giving rise to \textit{memory} in this hydrodynamic system. In the high-memory regime, walkers have been shown to mimic several hydrodynamic analogs of quantum systems. Some of these include orbital quantization in rotating frames \citep{Fort17515,harris_bush_2014,Oza2014} and confining potentials \citep{Perrard2014b,Perrard2014a,labousse2016,PhysRevE.103.053110}, Zeeman splitting in rotating frames \citep{Zeeman,spinstates2018}, wave-like statistical behavior in both confined geometries \citep{PhysRevE.88.011001,Giletconfined2016,Saenz2017,Cristea,durey_milewski_wang_2020} and open systems \citep{Friedal}, tunneling across submerged barriers \citep{Eddi2009,tunnelingnachbin,tunneling2020} and a macroscopic analog of spin systems~\citep{Saenz2021}. Walkers have also been predicted to show anomalous two-droplet correlations \citep{ValaniHOM,correlationnachbin}. Recently, efforts have also been made to develop a hydrodynamic quantum field theory for the walking-droplet system~\citep{Dagan2020hqft,Durey2020hqft}. A detailed review of hydrodynamic quantum analogs for walking droplets can be found in a recent review article by \citet{Bush2020review}.

%    \item Say what we are going to show in this letter briefly

In this Letter, we report anomalous transport behavior arising in a walking-droplet-inspired theoretical model that governs the dynamics of a one-dimensional wave-particle entity with a sinusoidal wave form in a tilted potential. We start by converting the integro-differential trajectory equation that governs the motion of the wave-particle entity into a system of Lorenz-like ordinary differential equations (ODEs) and perform a linear stability analysis to determine the stability of steady walking states. We then explore the different anomalous transport behaviors arising in both steady and unsteady walking regimes of the parameter space. %Finally, we comment on experimental setups that may be able to realize anomalous transport in the walking-droplet system and provide an outlook for future directions.

\begin{figure}
\centering
\includegraphics[width=\columnwidth]{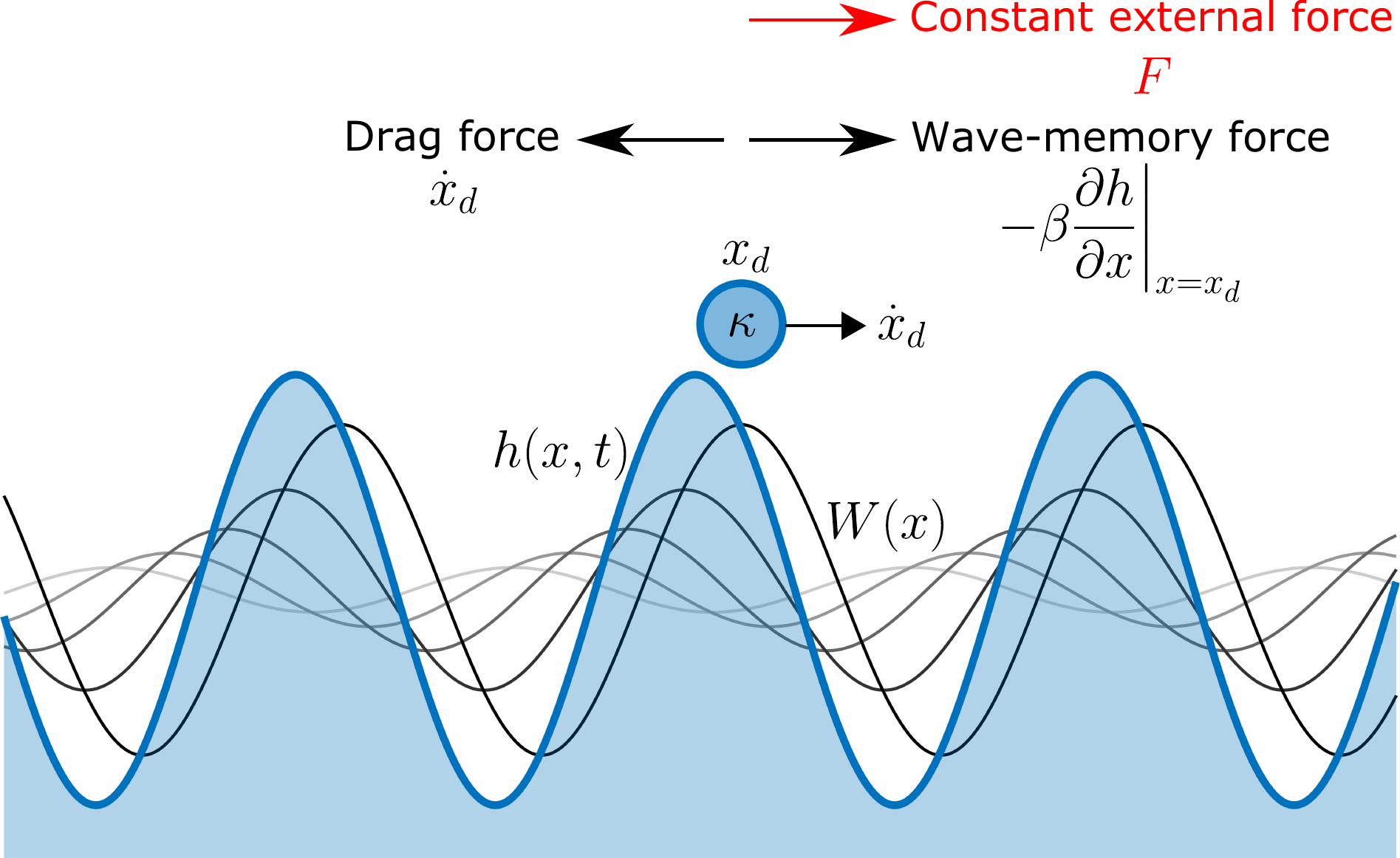}
\caption{Schematic of the one-dimensional self-propelled wave-particle entity. A particle of dimensionless mass $\kappa$ is located at $x_d$ and moving horizontally with velocity $\dot{x}_d$. The particle experiences a propulsion force,$-\beta\,\partial h/\partial x {|}_{x=x_d}$, from its self-generated wave field $h(x,t)$ (blue filled area), an effective drag force $-\dot{x}_d$ and a constant external force $F$. The underlying wave field $h(x,t)$ is a superposition of the individual waves continuously generated by the particle along its trajectory. These individual waves (black and gray curves with the higher intensity of the color indicating the waves created more recently) are of spatial form $W(x)=\cos(x)$, and decay exponentially in time.}
\label{Fig: schematic ANM}
\end{figure}

\begin{figure*}
\centering
\includegraphics[width=2\columnwidth]{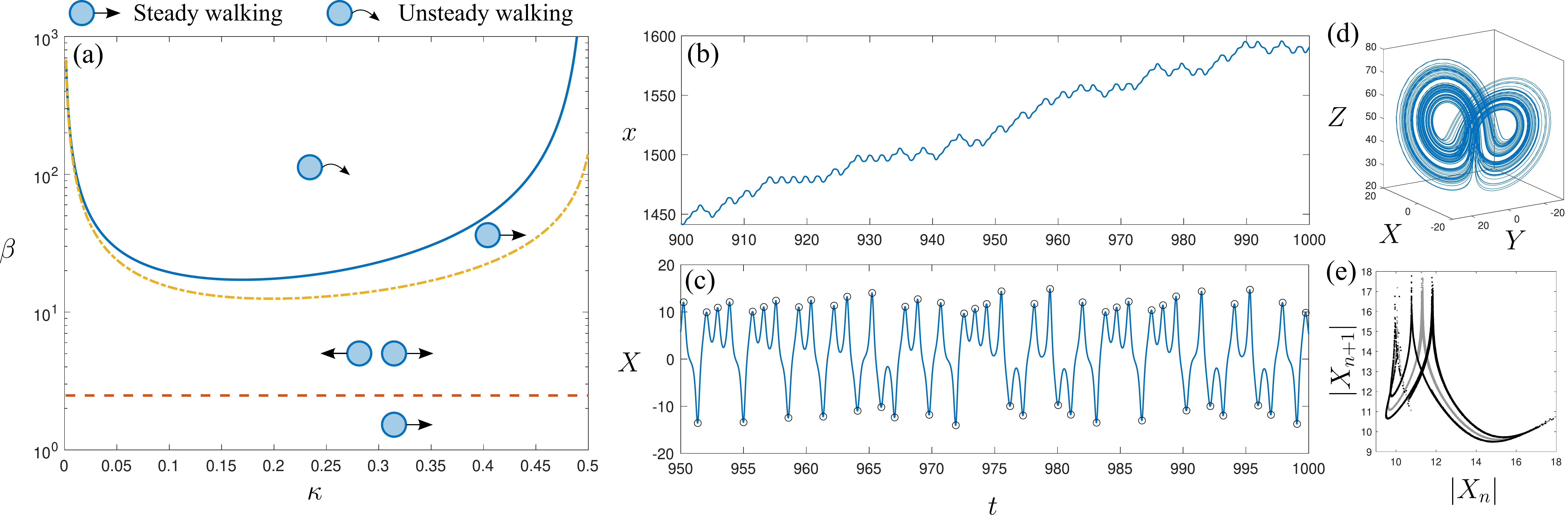}
\caption{(a) Linear stability diagram of the wave-particle entity in the $(\kappa,\beta)$ parameter space at $F=0.5$. Curves show parameter values when eigenvalues $\lambda$ of the dynamical system cross the $\text{Re}(\lambda)=0$ line, indicating a change in stability. Chaotic dynamics in the unsteady regime of the $(\kappa,\beta)$ parameter space for a typical parameter value of $\kappa=0.3$ and $\beta=50$ is shown in (b)-(d). (b) shows the space-time trajectory, (c) shows the corresponding velocity time series and (d) shows the underlying strange attractor that drives the chaotic dynamics. The black curve in (e) shows the $1$D return map of consecutive maxima (denoted by black circles in (c)) of absolute velocity $|X|$, while the gray curve shows the same for the classic Lorenz system with $F=0$.}
\label{Fig: LS ANM}
\end{figure*}

\textit{Theoretical model.} As shown schematically in Fig.~\ref{Fig: schematic ANM}, consider a particle located at position $x_d$ moving horizontally with velocity $\dot{x}_d$ while continuously generating waves with prescribed spatial structure $W(x)$ that decay exponentially in time. The dimensionless equation of motion governing the horizontal dynamics of the particle is given by~\cite{Oza2013}, 
\begin{equation}\label{eq: traj_1}
\kappa\ddot{x}_d+\dot{x}_d
=-\beta\frac{\partial h}{\partial x}\Big{|}_{x=x_d}+F.
\end{equation}
The left hand side of Eq.~\eqref{eq: traj_1} comprises an inertial term $\kappa\ddot{x}_d$ and an effective drag term $\dot{x}_d$, where the overdot denotes differentiation with respect to time $t$. The first term on the right hand side captures the forcing on the droplet by the underlying wave field $h(x,t)$. This force is proportional to the gradient of the underlying wave field. The second term is an external constant force $F$ arising from a tilted potential $V(x)=-Fx$. The shape of the wave field $h(x,t)$ is calculated through integration of the individual wave forms $W(x)$ that are continuously generated by the particle along its trajectory, giving
\begin{equation}\label{eq: traj_2}
h(x,t)=\int_{-\infty}^{t}W(x - x_d(s))\,\text{e}^{-(t-s)}\,\text{d}s.
\end{equation}
Combining Eqs.~\eqref{eq: traj_1} and \eqref{eq: traj_2}, one obtains the integro-differential equation
\begin{align}\label{eq_1}
\kappa\ddot{x}_d+\dot{x}_d
=\beta\int_{-\infty}^{t}f(x_d(t) - x_d(s))\,\text{e}^{-(t-s)}\,\text{d}s+F,
\end{align}
where $f(x)=-W'(x)$ is the negative gradient of the wave form and the prime denotes differentiation with respect to the argument $x$. The two parameters, $\kappa >0$ and $\beta >0$, follow directly from \citet{Oza2013} and may be usefully interpreted as the ratio of inertia to drag and the ratio of wave forcing to drag respectively. This integro-differential trajectory equation was derived by \citet{Oza2013} to describe the horizontal dynamics of a walking droplet by employing a Bessel function of the first kind and zeroth order, $W(x)=\text{J}_0(x)$, wave form for the individual waves generated by the droplet on each bounce. The model was recently investigated by \citet{Valani2021unsteady} and \citet{Durey2020lorenz} by employing a simpler sinusoidal wave form $W(x)=\cos(x)$ and it was shown by \citet{Valani2021unsteady} and \citet{ValaniLorenzlike2021} that the integro-differential equation of motion can be transform to the following set of ODEs (see Supplemental Material~\citep{supplement} for a derivation):
\begin{align} \label{Lorenz_droplet}
\dot{X}&=\sigma\left(Y-X+F\right), \\ \nonumber
\dot{Y}&=-XZ+rX-Y,\\ \nonumber
\dot{Z}&=XY-bZ.
\end{align}
These ODEs are the Lorenz equations with an added constant term $F$ in the first equation~\citep{Lorenz1963,Biasedlorenz}. Here, $X=\dot{x}_d$ is the droplet's velocity, $Y=\beta \int_{-\infty}^{t} \sin(x_d(t)-x_d(s))\,\text{e}^{-(t-s)}\,\text{d}s$ is the wave-memory force and $Z=\beta -\beta\int_{-\infty}^{t} \cos(x_d(t)-x_d(s))\,\text{e}^{-(t-s)}\,\text{d}s$ is also related to the wave-memory forcing. 
%The parameters $\sigma, r, b$ in the Lorenz system are related to the walker's system via $\sigma=1/\kappa$, $r=\beta/2$ and $b=1$.
Thus, the underdamped dynamics of an inertial particle of dimenionsionless mass $\kappa$ driven by a wave-memory force with coupling $\beta$, can alternatively be interpreted as the overdamped dynamics of a particle whose velocity $\dot{x}_d=X$ is driven by the Lorenz system with parameters $\sigma=1/\kappa$, $r=\beta$ and $b=1$. For the simulation results presented in this Letter, the system of ODEs in Eq.~\eqref{Lorenz_droplet} is solved in $\mathtt{MATLAB}$ using the inbuilt ODE solver ode45.

\textit{Steady solutions and linear stability analysis.} To obtain steady walking solutions of the wave-particle entity, we start by finding fixed points of the Lorenz-like system presented in Eq.~\eqref{Lorenz_droplet}. This gives the equilibrium solutions $X_0=u$, $Y_0=u-F$ and $Z_0=u(u-F)$, where the constant velocity $u$ satisfies the cubic equation~\citep{supplement}
\begin{equation}\label{eq: equil}
    u^3-F u^2 - (\beta-1) u - F=0.
\end{equation}
For $F > 0$ ($F<0$) and $0<\beta<1$, there is one real positive (negative) solution to Eq.~\eqref{eq: equil}, while for $\beta > 1$, one can conclude the existence of one real positive (negative) solution and either two or zero real negative (positive) solutions by invoking Descartes' rule of signs. To determine the stability of these steady walking solutions, one can perform a linear stability analysis by applying a small perturbation to the steady solutions~\citep{strogatz2019nonlinear}. This results in the following characteristic equation for the growth rate $\lambda$ of small perturbations~\citep{supplement}:
\begin{align*}
    & \kappa \lambda^3 + (2\kappa +1)\lambda^2 + \left[ \kappa (1+u^2) + 2 - \beta + u(u-F) \right] \lambda \\ \nonumber
    & + 1+3u^2 - \beta - 2uF=0.
\end{align*}
The linear stability diagram in the $(\kappa,\beta)$ parameter space for a fixed $F=0.5$ is shown in Fig.~\ref{Fig: LS ANM}(a). At small $\beta$ values, there is one stable steady walking solution corresponding to the wave-particle entity traveling in the direction of the applied force $F$. Above a critical value of $\beta$ (dashed red line) which is independent of $\kappa$, a stable-unstable pair of steady walking solutions emerges corresponding to the wave-particle entity walking steadily in the direction opposite to the applied force $F$. This results in a total of three steady walking solutions: a higher speed stable and a lower speed unstable solution in the direction opposite to $F$, and a stable solution in the same direction as $F$. Thus, in this region of multistability, we observe ANM with the wave-particle entity walking steadily in a direction opposite to the applied bias. At higher $\beta$ (above the yellow dotted-dashed curve), the stable steady walking solution in the direction opposite to $F$ becomes unstable, followed by the instability of the steady walking solution in the direction of $F$ (above the solid blue line). 

In the unsteady regime of the $(\kappa,\beta)$ parameter space at large $\beta$, we observe either chaotic dynamics or periodic oscillations in the particle's velocity $X$. Typical chaotic dynamics of the wave-particle entity driven by the underlying strange attractor in the unsteady regime is depicted in Fig.~\ref{Fig: LS ANM}(b)-(d). As shown in Fig.~\ref{Fig: LS ANM}(b), the particle seems to exhibits irregular diffusive-like behavior with a net drift in the direction of the force $F$. We note that in the absence of the force $F$, the particle has been demonstrated to exhibit diffusive-like behavior with no net drift~\citep{Valani2021unsteady}. The time series of the particle's velocity $X$ in Fig.~\ref{Fig: LS ANM}(c), the underlying strange attractor in the $(X,Y,Z)$ phase-space in Fig.~\ref{Fig: LS ANM}(d) and the $1$D return map of the maxima in $|X|$ in Fig.~\ref{Fig: LS ANM}(e), show similarities with the classic Lorenz system~\citep{Lorenz1963}. We note that a double-cusp structure (black) is observed in the $1$D return map here, as opposed to a single cusp structure (gray) observed for the classic Lorenz system. This single-cusp to double-cusp transition is due to the asymmetry introduced by the constant bias force $F$ which breaks the degeneracy of the left and right walking states. 

\begin{figure}
\centering
\includegraphics[width=\columnwidth]{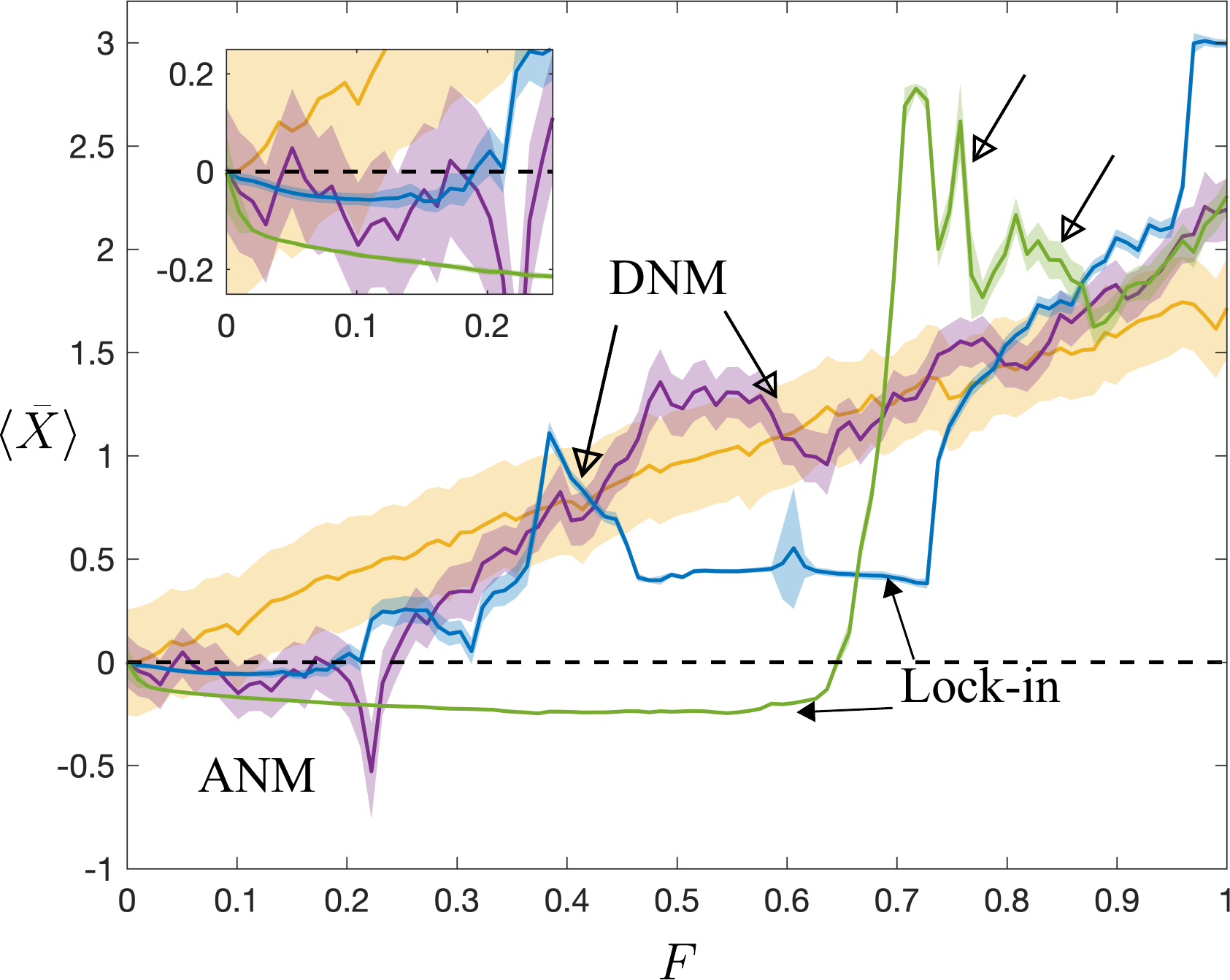}
\caption{Anomalous transport behavior. The average velocity $\langle \bar{X} \rangle$ of the particle against the applied constant force $F$ is shown for four sets of parameter values: $(\kappa,\beta)=(0.20,30)$ (yellow curve), $(0.17,67)$ (purple curve), $(0.25,100)$ (blue curve) and $(0.30,140)$ (green curve). The dashed horizontal line indicates $\langle \bar{X} \rangle=0$. The shaded region shows the standard deviation of $\bar{X}$ over the $1000$ simulated trajectories.}
\label{Fig: ANM fig}
\end{figure}

\begin{figure*}
\centering
\includegraphics[width=2\columnwidth]{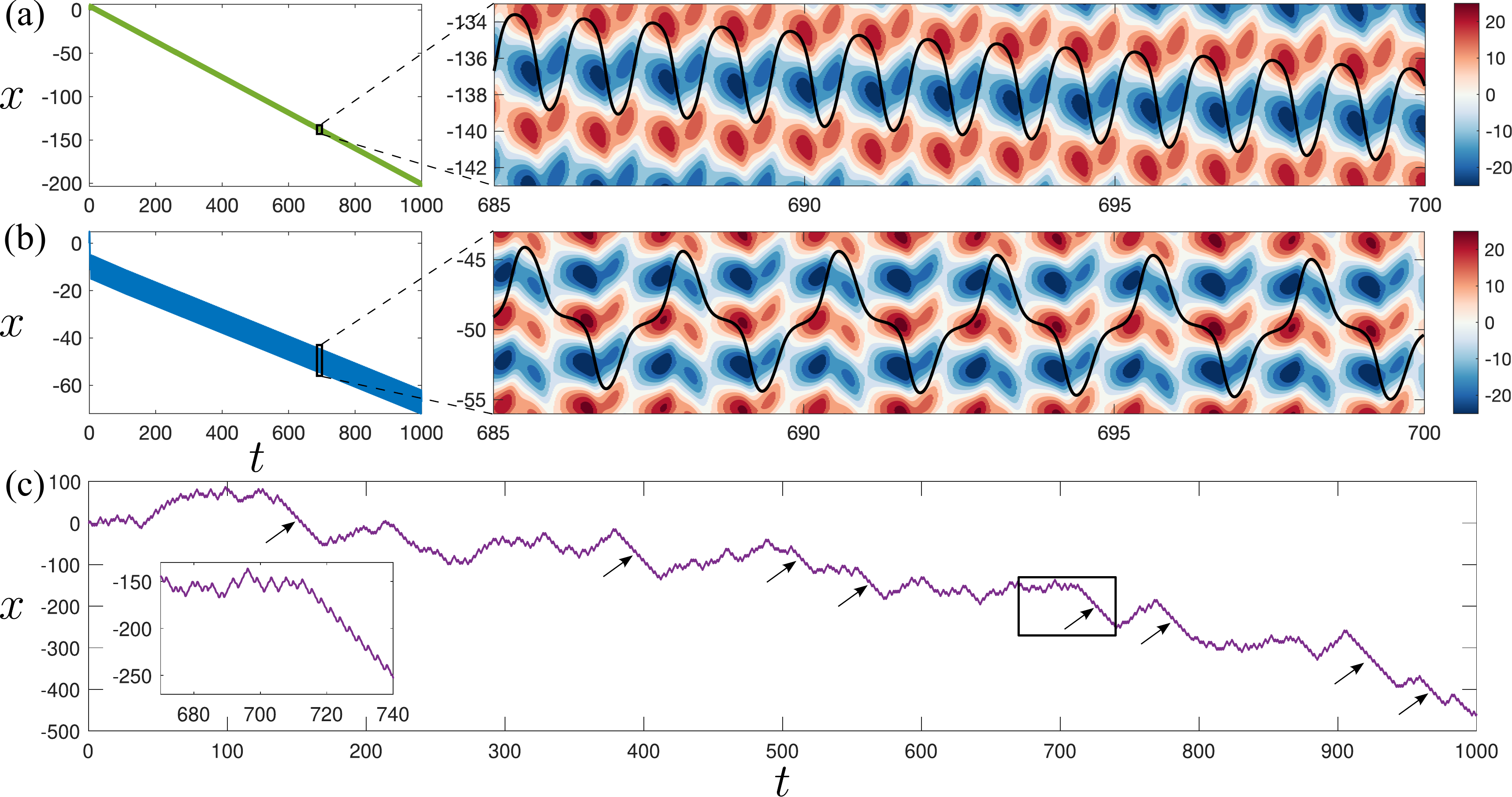}
\caption{Space-time trajectory of the particle during anomalous transport. Trajectory of the particle for ANM shown at (a) $(\kappa,\beta,F)=(0.30,140,0.2)$, (b) $(0.25,100,0.1)$ and (c) $(0.17,0.67,0.22)$. The enlarged trajectory in panels (a) and (b) also show contours of the underlying wave field $h(x,t)$. The arrows in panel (c) show bursts of periodic motion in the opposite direction to the applied force.}
\label{Fig: ANM traj}
\end{figure*}

\textit{Anomalous transport in the unsteady regime.} To investigate the particle's mobility in the unsteady regime, simulations were performed at fixed parameter values with $M=1000$ different random initial conditions. The initial values of the dynamical variables $X,Y$ and $Z$ were drawn randomly from a uniform probability distribution between $-1$ and $1$. A time-averaged particle velocity for a given initial condition was first calculated as $\bar{X}_j=\sum_{i=1}^{N} X_j(t_i)/N$, which was then ensemble averaged over all the initial conditions to give $\langle \bar{X} \rangle=\sum_{j=1}^{M} \bar{X}_j/M$. This average velocity $\langle \bar{X} \rangle$ as a function of the applied force $F$ is shown in Fig.~\ref{Fig: ANM fig} for four different sets of $(\kappa,\beta)$ values showing the various transport behaviors observed in the unsteady regime. For the parameter set $(\kappa,\beta)=(0.20,30)$ (yellow curve), we find that $\langle \bar{X} \rangle$ increases almost linearly in the direction of the applied force as the magnitude of the applied force increases. This indicates normal mobility behavior that one finds in equilibrium systems. For the parameter sets $(\kappa,\beta)=(0.17,67)$ (purple curve), $(0.25,100)$ (blue curve) and $(\kappa,\beta)=(0.30,140)$ (green curve), we clearly see regions of ANM where $\langle \bar{X} \rangle<0$. We have observed two qualitatively different types of ANM in the unsteady regime: (i) the particle is undergoing periodic back-and-forth oscillations with a net drift in the direction opposite to $F$ (see Fig.~\ref{Fig: ANM traj}(a)-(b)) and (ii) the particle's velocity is chaotic resulting in an irregular diffusive-like trajectory with a net drift in the direction opposite to $F$ (see Fig.~\ref{Fig: ANM traj}(c)). In the periodic ANM regime at parameter values $(\kappa,\beta,F)=(0.30,140,0.2)$, the particle oscillates back-and-forth between two consecutive peaks of the underlying wave field along with a net drift in the direction opposite to $F$, while, at parameter values $(\kappa,\beta,F)=(0.25,100,0.1)$, the particle is able to cross one peak during back-and-forth oscillations and hence oscillates with twice the wavelength along with a slower net drift opposite to $F$. In the chaotic ANM at parameter values $(\kappa,\beta,F)=(0.17,67,0.22)$, although the space-time trajectory exhibits irregular diffusive-like dynamics, we observe intermittent bursts of periodic drifts in the direction opposite to $F$ which results in an average motion opposite to the applied force. In addition to ANM, we also see in Fig.~\ref{Fig: ANM fig} regions of DNM which correspond to a negative slope in the curve when $\langle \bar{X} \rangle>0$, indicating that an increase in $F$ leads to decrease in $\langle \bar{X} \rangle$. We also encounter another anomalous transport behavior that we term lock-in regions. Here the average velocity $\langle \dot{X} \rangle$ stays almost constant for a range of $F$ values, indicating force-independent mobility. In these lock-in regions, the particle is typically undergoing periodic back-and-forth oscillations with a net drift either in the direction of $F$ or opposite to $F$.

\textit{Conclusions.} In this Letter, we have shown anomalous transport behaviors of a wave-memory driven particle under the presence of a constant external force. In the steady walking regime, ANM is observed in the multistable region where a stable steady walking solution is realized in the direction opposite to the applied force. In the unsteady walking regime, typically, the wave-particle entity undergoes either back-and-forth oscillations or irregular diffusive-like motion, with a net drift in the direction of the applied bias, but remarkably, regions of anomalous transport behaviors were observed where the wave-particle entity exhibits ANM, DNM and lock-in regions. %Although, we have used the idealized model of the wave-particle entity with a sinusoidal wave form for analytical tractability, ANM is also observed with a more experimentally accurate Bessel function wave form for the walking droplet (see Sec.~XX of Supplemental Material). 
The observations of anomalous transport reported in this Letter motivate experimental investigations with walking and superwalking droplets~\citep{superwalker,superwalkernumerical,ValaniSGM} in a tilted potential where these behaviors may be realized in experiments. Moreover, the present study also motivates both theoretical and experimental investigations of other counterintuitive non-equilibrium phenomena, such as ratcheting effects and stochastic resonance, that may arise in the walking-droplet system.  

\textit{Acknowledgements.} I would like to thank David M. Paganin for useful discussions.

\bibliography{ANM_WP}

\end{document}

\title{Supplemental Material for ``Anomalous transport of a classical wave-particle entity in a tilted potential''}% Force line breaks with \\
%\thanks{A footnote to the article title}%

\author{Rahil N. Valani$^{1}$}\email{rahil.valani@adelaide.edu.au}
\affiliation{$^1$School of Mathematical Sciences, University of Adelaide, South Australia 5005, Australia}

\maketitle

\section{Transforming the walker's integro-differential equation into a Lorenz-like system of ODEs}
Here we provide a derivation to convert the integro-differential equation of motion for the one-dimensional wave-particle entity into a Lorenz-like system of ODEs. The integro-differential equation governing the motion of the wave-particle entity in a tilted potential is given by
\begin{equation}\label{App: eq_1}
\kappa\ddot{x}_d+\dot{x}_d
={\beta}\int_{-\infty}^{t}\sin(x_d(t) - x_d(s))\,\text{e}^{-(t-s)}\,\text{d}s + F.
\end{equation}
Let us denote the wave-memory force by
\begin{equation*}
Y(t)={\beta}\int_{-\infty}^{t}\sin(x_d(t) - x_d(s))\,\text{e}^{-(t-s)}\,\text{d}s.
\end{equation*}
Differentiating $Y(t)$ with respect to time and using the Leibniz integration rule, we have
\begin{align}\label{App: eq_2}
\dot{Y}(t)=&-{\beta}\int_{-\infty}^{t}\sin(x_d(t) - x_d(s))\,\text{e}^{-(t-s)}\,\text{d}s\\ \nonumber
&+{\beta \dot{x}_d}\int_{-\infty}^{t}\cos(x_d(t) - x_d(s))\,\text{e}^{-(t-s)}\,\text{d}s\\ \nonumber
=&-Y(t)+X W(t),
\end{align}
where $\dot{x}_d=X$ and 
\begin{equation*}
W(t)={\beta}\int_{-\infty}^{t}\cos(x_d(t) - x_d(s))\,\text{e}^{-(t-s)}\,\text{d}s.        
\end{equation*}
Differentiating $W(t)$ with respect to time, we get
\begin{align}\label{App: eq_3}
\dot{W}(t)=&{\beta}-{\beta}\int_{-\infty}^{t}\cos(x_d(t) - x_d(s))\,\text{e}^{-(t-s)}\,\text{d}s\\ \nonumber
&- {\beta\dot{x}_d}\int_{-\infty}^{t}\sin(x_d(t) - x_d(s))\,\text{e}^{-(t-s)}\,\text{d}s\\ \nonumber
&={\beta}-W(t)-X Y(t).
\end{align}
By making the following change of variables
\begin{equation*}
Z(t)={\beta}-W(t), 
\end{equation*}
and substituting in Eqs.~\eqref{App: eq_2} and \eqref{App: eq_3} along with Eq.~\eqref{App: eq_1}, we obtain the following Lorenz-like system of ODEs~\citep{Lorenz1963,Biasedlorenz} for the dynamics of the wave-particle entity:
\begin{align}\label{eq: lorenzds}
    \dot{X}&=\frac{1}{\kappa}(Y-X+F)\\ \nonumber
    \dot{Y}&=-Y+{\beta}X-XZ\\ \nonumber
    \dot{Z}&=-Z+XY.\\ \nonumber
\end{align}

\section{Additional information on steady walking solutions and their linear stability analysis}

\subsection{Steady walking solutions}

To obtain steady walking solutions that correspond to the fixed points of the Lorenz-like system in Eq.~\eqref{eq: lorenzds}, we fix $(X,Y,Z)=(X_0,Y_0,Z_0)$ which results in $\dot{X}_0=\dot{Y}_0=\dot{Z}_0=0$ and gives
\begin{align*}
    0&=Y_0-X_0+F\\ \nonumber
    0&=-X_0Z_0+\beta X_0 - Y_0\\ \nonumber
    0&=X_0Y_0-Z_0.\\
\end{align*}
Solving these three equations simultaneously for $X_0,Y_0$ and $Z_0$ gives
\begin{equation*}
    X_0=u,\:\:Y_0=u-F\:\:\text{and}\:\:Z_0=u(u-F),
\end{equation*}
where $u$ is the constant steady walking velocity satisfying
\begin{equation*}
    u^3-F u^2 - (\beta-1) u - F=0.
\end{equation*}
This equation can be rearranged as follows:
\begin{equation}\label{eq: equil2}
u\left(u^2-(\beta-1) \right)=F(u^2+1).
\end{equation}
The left hand side of Eq.~\eqref{eq: equil2} is independent of the applied constant force $F$ and it geometrically represents a cubic in $u$, while the right hand side of Eq.~\eqref{eq: equil2} geometrically represents a parabola in $u$ which is dilated by a factor $F$ and has a vertical intercept at $F$. The intersection of these two curves determines the steady walking solutions of the wave-particle entity with a constant velocity $u$.

\begin{figure*}
\centering
\includegraphics[width=2\columnwidth]{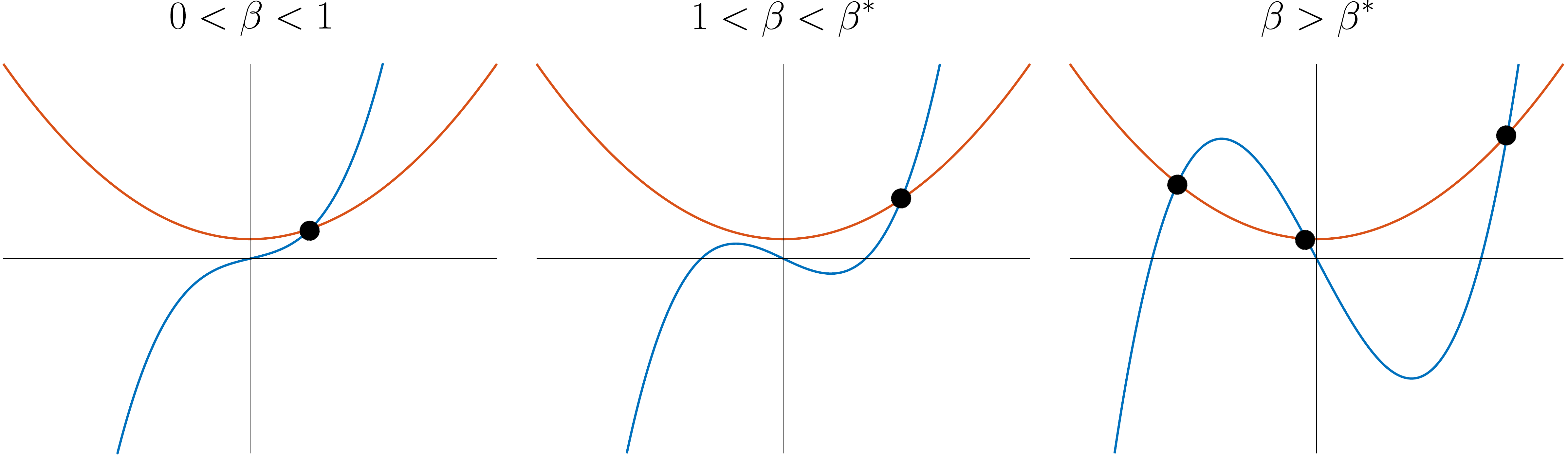}
\caption{Steady walking equilibrium solutions for $F>0$. For $0<\beta<1$, the cubic equation (blue) has an inflection point at the origin and results in a single intersection with the parabola (red) corresponding to a steady walking solution in the direction of $F$. For $1<\beta<\beta^*$, the cubic develops a local maximum in the left half plane but its vertical extent is not sufficient to intersect with the parabola, again resulting in a single steady walking solution in the direction of $F$. For $\beta>\beta^*$, we get three intersections between the cubic and parabola, giving us three steady walking solutions: two in the direction opposite to $F$ and one in the direction of $F$.}
\label{Fig:schematic ANM eq}
\end{figure*}
    
As shown schematically in Fig.~\ref{Fig:schematic ANM eq}, for $F>0$, when $0<\beta<1$, the cubic on the left hand side of Eq.~\eqref{eq: equil2} has a single horizontal intercept that corresponds to an inflection point at the origin. This results in a single intersection with the parabola in the right half of the plane, corresponding to a steady walking solution in the direction of $F$. For $1<\beta<\beta^*$, the cubic now has three horizontal intercepts at $0$ and $\pm\sqrt{\beta-1}$, however, the vertical extent of the local maximum in the left half plane is not sufficient to intersect with the parabola, again resulting in a single steady walking solution in the direction of $F$. For $\beta>\beta^*$, the vertical extent near the local maximum of the cubic in the left half plane is sufficient to intersect, resulting in two additional solutions. Thus, one now has three solutions: two steady walking solutions in the direction opposite to $F$ and one steady walking solution in the direction of $F$.

\subsection{Linear stability analysis}

To determine the stability of the steady walking solutions, we perform a linear stability analysis by perturbing the equilibrium solutions $(X_0,Y_0,Z_0)=(u,u-F,u(u-F))$. We add a small perturbation to the equilibrium solution as follows: $(X,Y,Z)=(u,u-F,u(u-F))+\epsilon(X_1,Y_1,Z_1)$, where $\epsilon>0$ is a small perturbation parameter. Substituting this in Eq.~\eqref{eq: lorenzds} and comparing terms of $\Big{O}(\epsilon)$ we get the following linear system that governs the evolution of perturbations,
\begin{gather}\label{eq: mateq}
 \begin{bmatrix} \dot{X}_1 \\
 \dot{Y}_1 \\
 \dot{Z}_1 
 \end{bmatrix}
 =
 \begin{bmatrix} -1/\kappa & 1/\kappa & 0 \\
\beta-u(u-F) & -1 & -u \\
u+B & u & -1
 \end{bmatrix}
  \begin{bmatrix} {X}_1 \\
 {Y}_1 \\
 {Z}_1 
 \end{bmatrix}.
\end{gather}
The solutions of Eq.~\eqref{eq: mateq} are proportional to $\text{e}^{\lambda t}$, with complex
growth rates $\lambda$ given by the eigenvalues of the right-hand-side matrix. The eigenvalues are obtained by solving the following determinant:
\begin{gather*}
\begin{vmatrix}
 -1/\kappa-\lambda & 1/\kappa & 0 \\
\beta-u(u-F) & -1-\lambda & -u \\
u+B & u & -1-\lambda
\end{vmatrix}=0.
\end{gather*}
This results in the following characteristic polynomial equation to be solved for the eigenvalues $\lambda$ which determine the growth rate of perturbations:
\begin{align*}
    & \kappa \lambda^3 + (2\kappa +1)\lambda^2 + \left[ \kappa (1+u^2) + 2 - \beta + u(u-F) \right] \lambda \\ \nonumber
    & + 1+3u^2 - \beta - 2uF=0.
\end{align*}

\bibliography{ANM_WP}